\documentclass{aa}
\usepackage{graphicx}
\usepackage{txfonts}

\begin{document}

  \title{Temperature and cooling age of the white dwarf companion of PSR~J0218+4232}

  \author{C. G. Bassa\inst{1}
    \and M. H. van Kerkwijk\inst{1}\thanks{\emph{Present address:} Dept. of Astronomy
      and Astrophysics, Univ. of Toronto, 60 St George Street, Toronto, ON M5S 3H8, Canada;
      \email{mhvk@astro.utoronto.ca}}
    \and S. R. Kulkarni\inst{2}}

  \institute{Astronomical Institute, Utrecht University, PO Box 80000, 3508 TA
    Utrecht, The Netherlands \and Palomar Observatory, California Institute of
    Technology 105-24, Pasadena, CA 91125, USA}

  \offprints{C.G. Bassa,\\ \email{c.g.bassa@astro.uu.nl}}

  \date{Received 7 February 2003 / Accepted 6 March 2003}

  \abstract{We report on Keck optical BVRI images and spectroscopy of
    the companion of the binary millisecond pulsar PSR~J0218+4232. A faint
    bluish ($V=24.2$, $B-V=0.25$) counterpart is observed at the pulsar
    location.  Spectra of this counterpart reveal Balmer lines which
    confirm that the companion is a Helium--core white dwarf. We find that
    the white dwarf has a temperature of
    $T_\mathrm{eff}=8060\pm150$~K. Unfortunately, the spectra are of
    insufficient quality to put a strong constraint on the surface
    gravity, although the best fit is for low $\log g$ and hence low mass
    ($\sim0.2\ \mathrm{M}_\odot$), as expected. We compare
    predicted white dwarf cooling ages with the characteristic age of the
    pulsar and find similar values for white dwarf masses of 0.19 to
    0.3~$\mathrm{M}_\odot$. These masses would imply a distance of
    2.5 to 4~kpc to the system. The spectroscopic observations also enable us to
    estimate the mass ratio between the white dwarf and the pulsar. We
    find $q=7.5~\pm~2.4$, which is consistent with the current knowledge
    of white dwarf companions to millisecond pulsars.

    }

  \maketitle

  \section{Introduction}
  Millisecond or recycled pulsars are almost always found in
  binaries. Their short (millisecond) periods are believed to be caused
  by the accretion of matter from the binary companion, which spins up
  the neutron star, and, by means poorly understood, decreases its
  magnetic field. This accretion will happen when the companion, in the
  course of its evolution, overfills its Roche lobe.

  When the companion of the pulsar is a white dwarf, one would expect
  the cooling age of the white dwarf to match the age of the
  pulsar. This because the cooling and spin-down clocks start
  ticking at roughly the same time, when the companion starts to
  contract to a white dwarf and the pulsar turns on following the
  cessation of mass transfer.

  The age of a spin-down powered pulsar, with period $P$ and period
  derivative $\dot{P}$, is given by
  \begin{equation}
    \tau_\mathrm{PSR} = \frac{P}{(n-1)\dot{P}}
    \left[ 1-\left(\frac{P_0}{P}\right)^{n-1} \right],
  \end{equation}
  where $P_0$ is the initial period and $n$ is the braking index. The
  latter determines how the rotational energy loss scales with the
  period ($\dot{P}\propto P^{2-n}$).  For magnetic dipole radiation (see
  Lyne \& Smith \cite{lyne}) we have $n=3$, and if we further assume
  that $P_0 \ll P$, the pulsar spin-down age is given by
  $\tau_\mathrm{char}=P/(2\dot{P})$, the ``characteristic age'' of the
  pulsar. The uncertainty in the pulsar age is caused by $n$ and $P_0$,
  where, for the former, values lower than 3 will lead to ages longer
  than the characteristic age, while initial periods close to the
  current period will lead to shorter ages.

  Another possible source of uncertainty in the characteristic pulsar
  age is the transverse velocity of the pulsar, which causes an
  apparent acceleration that contributes to the observed period
  derivative.  Ignoring this `Shklovskii' effect (Shklovskii
  \cite{shklovskii}) can lead to significant underestimates of the
  characteristic age (Camilo et al.\ \cite{camilo}).

  The cooling age of the white dwarf can be determined from its
  effective temperature and its mass, using a cooling model. With the
  optical detection of several white dwarf companions to millisecond
  pulsars (e.g., Danziger et al.\ \cite{danziger}; Nicastro et al.
  \cite{nicastro}; Lundgren et al.\ \cite{lundgren_a},b), considerable
  effort has been spent in constructing such cooling models (Hansen \&
  Phinney \cite{hansen_a}; Driebe et al.\  \cite{driebe_a}; Serenelli et
  al.\ \cite{serenelli}).  It was found that the main uncertainty in
  predicting white dwarf cooling ages is the thickness of the Hydrogen
  envelope surrounding the Helium--core. Thick envelopes can support
  residual Hydrogen shell burning, which slows the cooling of the white
  dwarf, while thin layers cannot and thus lead to faster cooling.

  Hydrogen shell burning can become unstable, leading to
  thermonuclear flashes, which may reduce the mass of the envelope until
  it cannot sustain Hydrogen fusion any longer. It was already found by
  Webbink (\cite{webbink}) that these thermal flashes do not occur for
  white dwarfs with masses below $\sim0.2\ \mathrm{M}_\odot$, a
  result that subsequently has been confirmed by Alberts et
  al.\ (\cite{alberts}), Driebe et al.\ (\cite{driebe_a}), Sarna et al.
  (\cite{sarna_a}), and Althaus et al.\ (\cite{althaus}).  This limit is
  due to the fact that flashes are caused by unstable Hydrogen
  burning through the CNO cycle (Driebe et al.\ \cite{driebe_b}) which
  does not occur for stars with masses below $\sim0.2\
  \mathrm{M}_\odot$.

  Above this limit all authors do find flashes, but the consequences
  differ.  Driebe et al.\ (\cite{driebe_a}) find that the flashes do not
  reduce the envelope mass sufficiently to suppress residual Hydrogen
  shell burning and hence they predict long cooling ages for masses also
  above $0.2\ \mathrm{M}_\odot$. In a later paper, however,
  Sch\"onberger et al.\ (\cite{schoenberger}) suggest this result
  might be invalid for closer binaries, in which the mass of the envelope
  might be reduced due to Roche-lobe overflow during a flash. Sarna et
  al.\ (\cite{sarna_b}) used binary evolution calculations to pursue
  this idea further. On the other hand, the Hydrogen envelope may also
  be reduced by nuclear burning during the flashes. Indeed, Serenelli
  et al.\ (\cite{serenelli}) and Rohrmann et al.\ (\cite{rohrmann_b})
  find that for masses above $\sim0.18\ \mathrm{M}_\odot$, their
  models experience one or more thermonuclear flashes, leading to 
  thinner Hydrogen envelopes that cannot sustain nuclear fusion and 
  hence to white dwarfs that cool much faster.

  The models are best compared with data.  In Table~\ref{bmsp}, we list
  the three millisecond pulsars with white dwarf companions for which
  the mass and temperature are known, and the cooling age can be
  determined. The first system, \object{PSR J1012+5307}, has a pulsar
  with a characteristic age of 8.9~Gyr, and a white dwarf with a
  temperature of $\sim8600$~K. The other two systems, 
  \object{PSR B1855+09} and \object{PSR J0437$-$4715}, both 
  have ages of 5~Gyr and white dwarf companions with 
  $T_\mathrm{eff}\sim4800$~K.

  For the cooling ages of the white dwarf companions of these systems to
  equal to the characteristic pulsar ages, one can readily see that
  there must a difference in the cooling properties of these white
  dwarfs. The systems PSR B1855+09 and PSR J0437$-$4715 are very similar
  to each other, but rather different from PSR J1012+5307, the latter
  being older, but, paradoxically having a hotter white dwarf
  companion. Given the cooling models and the masses of these white
  dwarfs, it seems likely that this
  discrepancy is due to a difference in thickness of the Hydrogen layer,
  those of PSR B1855+09 and PSR J0437$-$4715 being thin and that of PSR
  J1012+5307 being thick.

  Here we report on optical observations of the white dwarf companion of
  PSR~J0218+4232, which, as we will explain below, is expected to have a
  mass that is between the masses of PSR J1012+5307 and PSR
  J0437$-$4715, and thus, potentially, may help us gain insight in the
  cooling of white dwarfs. In Sect.~2, we outline our current knowledge
  of the PSR~J0218+4232 system. We describe our Keck observations in
  Sect.~3, from which, in Sect.~4, we determine the white dwarf
  temperature and surface gravity, together with radial--velocity
  measurements.  The mass of the white dwarf and the age and distance of
  the system are described in Sect.~5 and we discuss and conclude in
  Sect.~6.

  \section{PSR~J0218+4232}
  PSR~J0218+4232, a 2.3 ms pulsar, was discovered by Navarro et al.
  (\cite{navarro}). From radio timing observations, they found that it
  resides in a 2 day circular ($e<2\times10^{-5}$) orbit around a low
  mass object, with, assuming a pulsar mass of 1.4~$\mathrm{M}_\odot$, a
  minimum mass of $0.17\ \mathrm{M}_\odot$. Using the statistical
  argument that the probability of finding the orbit with an inclination
  $i$ less than $i_0$ is given by $P(i<i_0)=1-\cos i_0$, we can conclude
  that, with 90\% confidence, the mass of the companion is $0.17 \leq
  M_\mathrm{WD}/ \mathrm{M}_\odot \leq 0.44$, the range in which one
  would expect the object to be a white dwarf with a Helium--core.

  From considerations of the evolution of binaries containing
  millisecond pulsars with Helium-core white dwarf companions, one
  expects a relation between the mass of a Helium--core white dwarf
  companion to a millisecond pulsar and the orbital period (see Joss et
  al. \cite{joss}).  This relation has been modelled by Rappaport et al.\ (\cite{rappaport}),
  Tauris \& Savonije (\cite{tauris}) and others. 
  For the orbital periods of PSR~B1855+09 and PSR~J0437$-$4715 (Table~\ref{bmsp}),
  the relation by Tauris \& Savonije (\cite{tauris}) predicts companion masses to
  within the observed errors. Though this relation is not expected to be
  accurate for orbital periods less than 2 days, the predicted white
  dwarf mass of PSR J1012+5307, having an orbital period
  of 0.605 days, is within 1.5$\sigma$ of the observed result. Since the
  orbital period of PSR~J0218+4232 is in between that of PSR J1012+5307
  and that of PSR J0437$-$4715 (5.471 days) it is expected that
  PSR~J0218+4232 has a mass in between their masses. In fact, Tauris and
  Savonije (\cite{tauris}) predict a mass of $0.216\ \mathrm{M}_\odot$.
  This is near the mass of $\sim0.18\ \mathrm{M}_\odot$, where Althaus
  et al.\ (\cite{althaus}) predict the transition between thick and thin
  Hydrogen envelopes.

  From radio observations, a dispersion measure of 61.25 pc cm$^{-3}$
  was found. In the direction of the pulsar, the Taylor \& Cordes
  (\cite{taylor}) model for the distribution of interstellar free
  electrons predicts a minimum distance of 5.7 kpc for this dispersion
  measure. The quoted uncertainty is about 25\%, but may be considerable
  larger in this direction as there are few pulsars with which the model
  could be calibrated. Indeed, a new model, presented by Cordes \& Lazio
  (\cite{cordes}), predicts a distance of 2.7 kpc.

  The characteristic age of the pulsar is 0.46 Gyr (Navarro et al.\
  \cite{navarro}).  Recent radio observations of PSR~J0218+4232 with
  WSRT/PuMa (B. Stappers, priv.\ comm.) have reveiled a proper motion,
  of $\mu~=~3.97~\pm~1.73$~mas/yr.  Even for the largest distances we
  will need to consider, this has negligible effect on the
  characteristic age, and hence we ignore it below. 

  \section{Observations}
  The location of the PSR~J0218+4232 system was observed with the 10-m
  Keck I telescope at Hawaii, using the low--resolution imaging
  spectrometer (LRIS, Oke et al.\ \cite{oke}). During the night of 1995
  November 21/22, R and I--band images were obtained. The following night
  single deep B, V and R images were taken. In both nights the seeing
  was mediocre, 1\farcs0--1\farcs1, and the conditions were not
  photometric. Indeed, the images from the first night were of so much
  inferior quality than those taken the other night that we did not
  use them any further.

  On the night of 1997 January 8/9 the conditions were photometric, and
  short exposures of the system were taken in B, V, R and I-band for
  photometric calibration. The seeing was again mediocre and varied
  between 1\farcs0--1\farcs2. Four standard fields (Landolt
  \cite{landolt}; Stetson \cite{stetson_b}) were also observed in the
  same bands. Table~\ref{imalog} shows the log of the imaging
  observations.

  Spectra of the companion of PSR~J0218+4232 were obtained with the same
  instrument, now at Keck II, on the nights of 1998 December 16/17 and
  17/18. The 600 lines mm$^{-1}$ grating was used, covering the
  wavelength range 3500-6000~\AA~at 1.25~\AA~pix$^{-1}$. The width of
  the slit was 1\arcsec, resulting in a resolution of $\sim$6~\AA. The
  seeing varied between 0\farcs8--1\farcs0 on both nights.  

  On both nights that the pulsar companion was observed, the white
  dwarfs \object{WD 0030+444} and \object{WD 0518+333} (Reid
  \cite{reid}) were observed by way of radial--velocity standards, and the
  spectrophotometric standard \object{G 191B2B} (Bohlin et al.
  \cite{bohlin}) was observed for flux calibration. Exposures of a
  HgKrXe lamp were taken for wavelength calibration.  All spectra were
  taken at a position angle close to the parallactic one, to minimize
  loss of light and velocity shifts due to differential refraction.  For
  PSR~J0218+4232, the angle was chosen such that there would be another
  nearby star in the slit.  The spectroscopic observations are tabulated
  in Table~\ref{speclog}.

  \begin{table}
    \caption[]{Log of the Keck imaging observations.}\label{imalog}
    \begin{tabular}{lccr}
      \hline
      Object$^\mathrm{a}$ & Time (UT) & $t_\mathrm{int}$ (s) & $\sec z$ \\
      \hline
      \multicolumn{2}{l}{1995 November 23}  & B,V,R & \\
      PSR J0218 & 7:24--7:56 & 1200,900,600 & 1.09--1.12 \\[0.8ex]
      \multicolumn{2}{l}{1997 January 9} & B,V,R,I & \\
      PSR J0218 & 5:47--5:56 & 120,60,60,60 & 1.09 \\
      PG 0231    & 5:18--5:33 & 2$\times$(20,10,10,10) & 1.03--1.04 \\
      SA 95      & 5:36--5:43 & 10,10,10,10 & 1.13 \\
      SA 95      & 7:35--7:42 & 10,10,10,10 & 1.08\\
      PG 104    & 15:40--15:53 & 2$\times$(20,10,10,10) & 1.19--1.22 \\
      PG 132    & 15:56--16:02 & 20,10,10,10 & 1.15 \\
      \hline
    \end{tabular}
    \begin{list}{}{}
    \item[$^\mathrm{a}$] Full names are PSR~J0218+4232,
      \object{PG 0231+051},\\ \object{PG 1047+003} and \object{PG 1323$-$086}.
    \end{list}
  \end{table}

  \begin{table}
    \caption[]{Log of the Keck spectroscopic observations.}\label{speclog}
    \begin{tabular}{lccr}
      \hline
      Object & Time (UT) & $t_\mathrm{int}$ (s) & $\sec z$ \\
      \hline
      \multicolumn{4}{l}{1998 December 17} \\
      WD 0030+444    & 5:47       & 180           & 1.11 \\
      PSR~J0218+4232 & 6:05--9:16 & 4$\times$2700 & 1.08--1.26 \\
      WD 0518+333    & 9:24       & 180           & 1.04 \\
      G 191B2B       & 9:36       & 30            & 1.19 \\[0.8ex]
      \multicolumn{4}{l}{1998 December 18} \\
      WD 0030+444    & 5:00       & 180           & 1.10 \\
      PSR~J0218+4232 & 5:11--8:24 & 4$\times$2700 & 1.08--1.17 \\
      WD 0518+333    & 8:34       & 180           & 1.08 \\
      G 191B2B       & 8:48       & 30            & 1.21 \\
      \hline
    \end{tabular}
  \end{table}

  \subsection{Astrometry}
  For astrometric calibration, we selected all 85 stars from the
  USNO--A2.0 Catalog (Monet et al. \cite{monet}) that overlapped with a
  10 second R-band image obtained on November 23, 1995. Of these stars,
  57 were not over exposed and appeared stellar and unblended. Their
  centroids were measured and corrected for instrumental distortion
  using a bicubic function determined by J. Cohen (1997, priv.
  comm.). These points were fitted for zero-point position,
  scale and position angle against the USNO--A2.0 positions. After
  rejecting 4 outliers, that had residuals larger than 0\farcs6, the rms
  residuals were 0\farcs18 and 0\farcs22 in right ascension and
  declination respectively, consistent with the expected errors for
  USNO-A2.0 positions.

  This solution was transferred to the 10 minute R-band (and the 20
  minute B-band) image, using 184 (209) stars that appeared on both
  images and were stellar, unsaturated and not blended. Again the
  zero-point, scale and position angle were fitted and the final
  residuals were $<$0\farcs03 ($<$0\farcs06) in both directions.
  Thus, our positions should be on the USNO--A2.0 system to within
  0\farcs03.  The USNO-A2.0 system is on the International Celestial
  Reference System (ICRS) to within about 0\farcs2.

\begin{figure*}
    \centering
    \includegraphics[width=17cm]{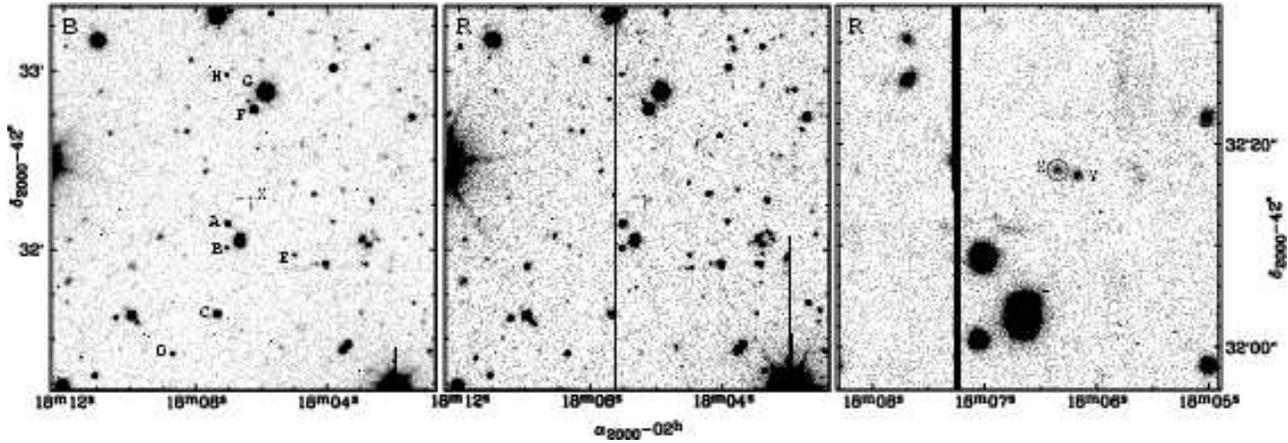}
    \caption{Optical images of the location of PSR~J0218+4232 and its
      companion.  The left-hand panel shows the 1200 s B image, the
      middle panel the 600 s R-band image, and the right-hand panel
      a zoom-in on the R-band image.  In the latter, the circle
      indicates the timing position of Navarro et al.\
      (\cite{navarro}).  For clarity, it has been drawn with a
      radius of 1\arcsec, twice the 95\% confidence error radius.
      We find one object within the circle and name it object X.
      Other stars in this field are also named and magnitudes of
      these stars are given in Table~\ref{magnitudes}.  Object Y,
      clearly visible in R but very faint in B, is only 2\arcsec
      west of the pulsar location.}
    \label{pulsarloc}
  \end{figure*}

  This latter uncertainty is much larger than the measurement error for
  our optical position, and also greatly exceeds the precision with
  which the very precise timing position (Navarro et al.\
  \cite{navarro}; B. Stappers, 2003, priv.\ comm.) can be tied to the
  ICRS.  Within this uncertainty, our images show a single object, `X'
  hereafter, at a position coincident with the timing position (see
  Fig.~\ref{pulsarloc}).  This object is also present on the 1997
  images.  

  \subsection{Photometry}
  We used the DAOPHOT II package (Stetson \cite{stetson_a}) for the
  photometry.  First, all images were bias-subtracted and flat-fielded
  using twilight flats.  Next, we determined a point spread function
  (PSF) for each object image from a selection of relatively isolated
  and single stars. We determined instrumental magnitudes by fitting
  this PSF to all stars. These fitting results were then used to remove
  all stars, except those used in determining the PSF. Using aperture
  photometry we determined the magnitudes of these PSF stars and
  calculated offsets between the PSF magnitude and the aperture
  magnitude.

  Instrumental magnitudes of the standards stars were determined using
  aperture magnitudes. This was necessary since the images of the
  standard fields were deliberately taken out of focus to ensure that
  the stars would not saturate the detector. Using values of Stetson
  (\cite{stetson_b}) we determined magnitude offsets, colour and
  extinction coefficients to calibrate the 1997 images. The rms
  residuals after the calibration are smaller than 0.02 magnitudes.

  Finally, magnitude offsets between the 1995 and 1997 images were
  determined to transfer the photometric calibration to the 1995 images. We find
  that for B, V and R respectively 90\%, 76\% and 85\% of the flux
  actually reached the detector, consistent with the presence of thin
  cirrus during the 1995 observations. The magnitudes of the pulsar
  companion and other stars in the vicinity are tabulated in
  Table~\ref{magnitudes}. We estimate the final uncertainty in the zero
  points at about 0.02 magnitudes.

  \begin{table}
    \caption[]{Astrometry and photometry of the companion of
      PSR~J0218+4232 and stars in the field. The nomenclature of the stars
      is according to Fig.~\ref{pulsarloc}. The uncertainties for star X,
      the pulsar companion, are as listed.  For the other
      stars, the positions are good to within 0\farcs01 except for those
      marked with a colon, which are good to within 0\farcs02. For the
      photometry of the other stars, the presence of no, a single or a
      double colon indicates uncertainties of $\sigma \leq 0.025$,
      $0.025 < \sigma \leq 0.075$ and $\sigma > 0.075\,$mag,
      respectively. These errors are 
      instrumental, i.e., they do not include the zero-point uncertainty of
      the astrometric tie (0\farcs2 in each coordinate) or of 0.02 mag in
      the photometry. The B, V and R magnitudes are determined from the 1995
      images while the I-band magnitudes are found from the 60~s image taken
      in 1997, hence the larger I-band error for object X.}\label{magnitudes}
    \begin{tabular}{lllllll}
      \hline
      & \multicolumn{2}{c}{$\alpha$ (J2000) $\delta$} & & & & \\
      ID & $2^\mathrm{h}18^\mathrm{m}$ & $+42\degr$ & $B$ & $V$ & $R$ & $I$ \\
      \hline
      X & 06\fs364 & 32\arcmin17\farcs40 &     24.40   &     24.15   &     23.86   &     23.76   \\
      & $\pm$0\farcs03 & \phantom{00}$\pm$0\farcs03 & \phantom{0}\llap{$\pm$}0.04   & \phantom{0}\llap{$\pm$}0.04
      & \phantom{0}\llap{$\pm$}0.06   & \phantom{0}\llap{$\pm$}0.25   \\[0.8ex]

      A & 07\fs029 & 32\arcmin08\farcs81 &     21.79   &     20.35   &     19.43   &     18.65   \\
      B & 07\fs066 & 32\arcmin00\farcs79 &     23.43   &     21.88   &     20.86   &     19.77   \\
      C & 07\fs353 & 31\arcmin38\farcs67 &     20.73   &     20.00   &     19.55   &     19.18   \\
      D & 08\fs700 & 31\arcmin25\farcs35 &     23.09   &     21.60   &     20.62   &     19.67   \\
      E & 05\fs015: & 31\arcmin58\farcs27 &     23.46   &     22.55   &     21.98   &     21.44:  \\
      F & 06\fs238 & 32\arcmin47\farcs09 &     20.64   &     19.02   &     17.88   &     16.58   \\
      G & 05\fs868 & 32\arcmin52\farcs90: &     17.23   &     16.81   &     16.42:  &     16.15   \\
      H & 07\fs058 & 32\arcmin58\farcs75 &     23.30   &     22.69   &     22.18   &     21.41:  \\
      Y & 06\fs194: & 32\arcmin17\farcs00: &     26.37:: &     24.78:: &     23.67:  &     22.19:  \\
      \hline
    \end{tabular}
  \end{table}

  \subsection{Spectroscopy}
  \label{spectroscopy}
  All frames were bias-corrected and sky-subtracted using standard
  procedures.  The spectra were extracted using an optimal extraction
  method similar to that presented by Horne (\cite{horne}).  Each
  extracted spectrum was wavelength calibrated using the arc lamp
  exposures, with a correction as detailed below.  Then, each individual
  spectrum was flux calibrated using the observation of G 191B2B of that
  night.

\begin{figure*}
    \centering
    \includegraphics[angle=270,width=17cm]{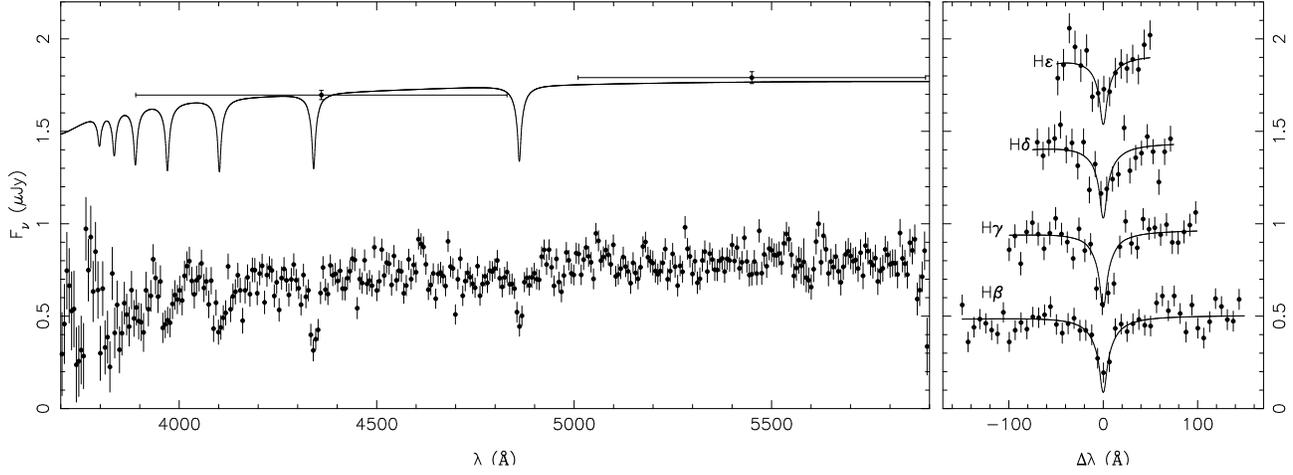}
    \caption{The spectrum of the companion of PSR~J0218+4232. Shown in
      the left-hand panel is the average of the observed spectra when
      shifted to zero-velocity (bottom curve) and the convolved
      modelled spectrum with $T_\mathrm{eff}=8000$~K and $\log g=7.0$
      cm s$^{-2}$ (top curve), shifted 1 $\mu$Jy upwards. Superposed
      to the latter are the observed B and V broadband fluxes (also offset
      by $1\,\mu$Jy). The
      right-hand panel shows the H$\beta$ to H$\epsilon$ lines from
      the observed spectrum with the convolved model superposed.}
    \label{spectrum}
  \end{figure*}

  The spectra of PSR~J0218+4232 were extracted using the spatial profile
  determined from the brighter star that we had made sure was on the
  slit.  After extraction, flux and wavelength calibration, all four
  spectra of each night were added and averaged.  The average of the two
  averaged spectra is shown in Fig.~\ref{spectrum}.  The Balmer lines
  H$\beta$, H$\gamma$ and H$\delta$ are clearly visible, indicating a
  Hydrogen atmosphere, in line with the object being a Helium--core white
  dwarf.

  As we hoped to determine a radial-velocity orbit, we took particular
  pains in trying to verify our wavelength calibration and velocity
  scale.  In the process, we encountered a number of problems, which we
  detail below.  As will become clear in the remainder of this Section,
  we have not solved the problems, and hence there is an additional
  uncertainty in the velocities we derive.  While in all likelihood this
  is smaller than the rather large measurement errors for
  PSR~J0218+4232, we still will treat our velocities with caution.  We
  stress, though, that the problems we encountered should have no effect
  on the main results of this paper, which rely on the temperature
  determination from the spectra.

  As a first correction to the wavelength calibration, we measured the
  wavelength difference between the observed and known wavelength of the
  \ion{O}{i} $\lambda5577$ night-sky emission line.  We found that these
  differences had a rms of 0.7~\AA.  We corrected all spectra
  appropriately.

  Next, we determined radial velocities of the two radial-velocity
  standards for each night by fitting a 9 parameter profile to each
  line, H$\beta$ up to H9, separately. This profile consisted of a
  parabolic term, added to the sum of a Lorentzian and a Gaussian, both
  having the same centre, but different widths and amplitudes. This
  profile fitted the lines extremely well, but unfortunately we found
  that the resulting velocities for the different lines were
  inconsistent with each other.

  After correcting for the barycentric motion of the Earth, the velocity
  determined from the H$\beta$ line in both spectra of WD 0518+333 was
  consistent with the results from Reid (\cite{reid}), while the other
  lines gave higher velocities. In general, the H$\gamma$, H$\delta$ and
  H$\epsilon$ lines gave velocities that were approximately 56, 78 and
  80 km~s$^{-1}$ too red.  The velocities determined from H8 and H9 had
  too large an error to be meaningful.

  For the spectrum of WD 0030+444 taken on the first night, we found a
  similar result, with the H$\beta$ velocity in agreement with the known
  value and the velocities of other lines following the pattern
  described above. The results for the second night, however, differed
  from those of the first night. First, the H$\beta$ velocity was slow by
  25 km~s$^{-1}$, and second, the other velocities followed a different
  pattern. 

  To verify our methods, we used the flux calibration standard G 191B2B,
  which is a hot sub-dwarf and which also has the H$\beta$ to
  H$\epsilon$ lines in its spectra, though they are much less wide than
  those of WD 0518+333 and WD 0030+444.  Its radial velocity has not
  been determined as accurately, but we can still use it for relative
  calibration.  We determined velocities with the same method as
  described above, and found that these followed the same pattern as the
  observations of WD 0518+333 and the first observation of WD 0030+444.

  On closer inspection, we found that the observation of WD 0030+444 on
  the second night differed in flux from the observation on the first
  night, the second spectrum having 40\% fewer counts. Using the filter
  curve from Bessell (\cite{bessell}), we calculated synthetic B-band
  magnitudes for both spectra.  We found that the first spectrum gave a
  magnitude consistent with the literature value, while the second
  spectrum obviously did not. Since there was no difference in airmass and
  conditions, we believe that WD 0030+444 was incorrectly placed on the
  slit, thus causing the inconsistent velocities and the flux
  difference.  Therefore, we decided not use the observation of WD
  0030+444 on the second night any further.

  We tried to find a reason for the trend in velocity found in the
  different Balmer lines, but could not find any.  Therefore, we simply
  used the velocities found from H$\beta$, H$\gamma$, H$\delta$ and
  H$\epsilon$ (H8 and H9 had too large an error) to calculated a new
  empirical wavelength scale, depending linearly on the wavelength
  calibration found from the arc lamp measurements.  This correction was
  applied to all spectra, after applying the shift for the wavelength difference
  found from the night-sky line.  

  With both corrections in place, the resulting velocities (see
  Table~\ref{wdradvel}) have root-mean-square errors of approximately 14
  km~s$^{-1}$.  We take this as an estimate of the additional
  uncertainty in any velocity measurement, but stress that since we do
  not know the cause of the problem with the wavelength calibration, we
  cannot be certain it applies to all observations.  As noted, however,
  it should not influence our main results, which rely on the
  temperature determination.

  \begin{table}
    \caption[]{Radial velocity measurements of the radial velocity
      standards WD 0030+444 and WD 0518+333 and the spectrophotometric
      standard G 191B2B.}\label{wdradvel}
    \begin{tabular}{cccc}
      \hline
      Date (UT) & \multicolumn{3}{c}{$v$ (km~s$^{-1}$)} \\
      & WD 0030+444 & WD 0518+333 & G191B2B \\
      \hline
      1998 Dec.\ 17 & $78.8\pm3.2$ & $34.7\pm2.3$ & $22.3\pm4.0$\\
      1998 Dec.\ 18 & $35.1\pm4.4$\rlap{$^\mathrm{a}$} & $34.7\pm2.6$ & $\phantom{0}9.9\pm4.3$\\
      \hline
    \end{tabular}
    \begin{list}{}{}
    \item[$^\mathrm{a}$]We believe this radial velocity
      measurement is in error due to misplacement of this object
      onto the slit.
    \end{list}
  \end{table}

  \section{Temperature, gravity and radial velocities}

  We use the photometry and the spectra of the companion of
  PSR~J0218+4232 to constrain its effective temperature and surface
  gravity by comparing them with model atmospheres.  Furthermore, we
  determine radial velocities of the two spectra and use these to
  estimate the mass ratio between the pulsar and the white dwarf and the
  systemic velocity of the system.

  We take into account the reddening in the direction of the pulsar.
  Schlegel et al (\cite{schlegel}) give $E_{B-V}=0.07\pm0.01$ and we
  correct our magnitudes accordingly. This reddening value was also used
  to correct the observed spectra using relations given by O'Donnell
  (\cite{odonnell}).

  The two separate average spectra of the companion, taken on the two
  subsequent days, were compared with model spectra for pure Hydrogen
  atmospheres, kindly provided by P. Bergeron (2002, priv.
  comm.), to infer the effective temperature and the surface
  gravity. The grid of models used spanned $T_\mathrm{eff}=$ 7000 (500)
  9500 K and $\log g=$ 6.5 (0.5) 8.0 cm s$^{-2}$, where the number in
  parentheses is the step between the models. We convolved each model
  with a cut-off Gaussian to mimic the effects of seeing and the slit.
  This convolved model was then fitted to each observed spectrum, using data-points
  around H$\beta$ up to H$\epsilon$, leaving the radial velocity $v$,
  flux normalisation $A$ and slope $\alpha$ to be fitted by minimising
  the $\chi^2$ merit function, defined by
  \begin{equation}
    \chi^2=\sum_i \left(\frac{f_i-f[(1+v/c)\lambda_i]\times A
      (\lambda_i/4340\ \mathrm{\AA})^\alpha}{\sigma_i}\right)^2.
  \end{equation}
  Here $f_i$ and $\sigma_i$ are the observed flux and error at
  wavelength $\lambda_i$, and $f[\lambda]$ is the model flux at
  wavelength $\lambda$. We introduce the slope $\alpha$ to account for
  possible errors in the (relative) flux calibration.

\begin{figure}
    \resizebox{\hsize}{!}{\includegraphics[angle=270]{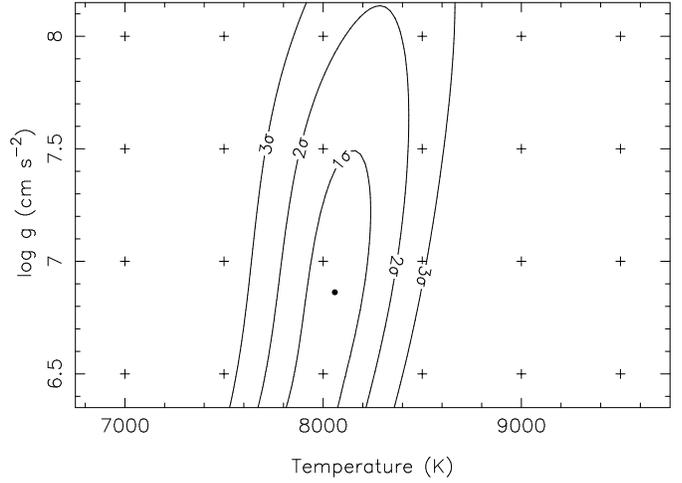}}
    \caption{Confidence intervals on effective temperature
      $T_\mathrm{eff}$ and surface gravity $\log g$ as inferred from our
      spectra.  The plus signs indicate the parameter combinations of
      the models in our grid.  At each point, the best-fit $\chi^2$ was
      determined, leaving velocity, flux amplitude, and spectral slope
      free to vary.  The results were interpolated to create a $\chi^2$
      surface, which has a minimum at the position shown with a filled
      circle, at $T_\mathrm{eff}=8060\pm150$ K and $\log g=6.9\pm0.7$ cm
      s$^{-2}$.  The three contours, at $\chi^2_\mathrm{min}$ plus 1, 4
      and 9, delineate the 1$\sigma$, 2$\sigma$ and 3$\sigma$ error
      regions.}
    \label{tefflogg}
  \end{figure}

  For each of the two observed spectra a surface of $\chi^2$ values in
  the $T_\mathrm{eff}$, $\log g$ plane, spanned by the modelled spectra,
  was created.  For each model spectrum the $\chi^2$ value of the first
  spectrum was added to that of the second to create a third surface.
  In Fig.~\ref{tefflogg}, we show contours and the location of the
  minimum in this latter surface; the surfaces found from the individual
  spectra are similar in shape. At the minimum, we find $\chi^2_\mathrm{min}=778.8$. 
  As we have 606 degrees of freedom, this indicates that the fit is 
  satisfactory, but that likely we underestimated our errors slightly.
  The errors in temperature and surface gravity were
  rescaled accordingly. The best-fit power-law slopes for the two spectra
  are small, $-0.11\pm0.16$ and $0.21\pm0.19$ for the first and second night,
  respectively, indicating that our relative flux calibration was reasonably
  good.  Indeed, also the absolute calibration is fair: the B-band
  magnitude, using the filter curve by Bessell
  (\cite{bessell}),  from our observed spectra is 24.43, within the
  error of the photometric measurement.

  From Fig.~\ref{tefflogg}, one sees that our spectra constrain the
  temperature fairly well ($T_\mathrm{eff}=8060\pm150$ K), but provide
  only a very weak constraint on the gravity ($\log g=6.9\pm0.7$ cm
  s$^{-2}$).  This is because the gravity mostly affects the higher
  Balmer lines, at wavelengths shortwards of 4000~\AA, where our spectra
  are very noisy.

  The photometric colours of the companion can also be used to infer a
  temperature.  From atmosphere models by Bergeron et al.\
  (\cite{bergeron_a}), we find a temperature of
  $T_\mathrm{eff}=8560\pm700$ K for a pure Hydrogen atmosphere with
  $\log g=8$. Serenelli et al.\ (\cite{serenelli}) and Rohrmann et al.\
  (\cite{rohrmann_b}) calculated evolutionary tracks of the formation
  and cooling of Helium white dwarfs. They also give colours based on
  model atmospheres of Rohrmann (\cite{rohrmann_a}). From these we infer
  a temperature of $T_\mathrm{eff}=8400\pm500$ K over a range of surface
  gravities ($\log g=6.0-7.5$). Both these temperatures are consistent
  with what we found from our spectra.

  The radial velocities that we found from fitting the spectra to the models by
  Bergeron are given in Table~\ref{psrj0218radvel}.  These velocities are
  corrected for the barycentric motion of the earth and are for the best fit of
  $T_\mathrm{eff}$ and $\log g$. The error estimates include the extra
  uncertainty in each velocity due to the uncertainties in the temperature,
  gravity, normalisation, and slope.  

  \begin{table}
    \caption[]{Radial velocity measurements of the companion of
      PSR~J0218+4232.  The uncertainties listed are measurement errors,
      and do not take into account the possible systematic errors
      associated with the problems in the wavelength calibration
      (Sect.~\ref{spectroscopy}).} 
    \label{psrj0218radvel}
    \begin{tabular}{ccccr}
      \hline
      \multicolumn{2}{c}{Date \& Time (UT)} & MJD$_\mathrm{bar}$ & $\phi_\mathrm{orb}^a$ & $v$ (km~s$^{-1}$) \\
      \hline
      1998 Dec.\ 17 & 07:45 & 51164.3281 & 0.544 & $-211 \pm 64$ \\
      1998 Dec.\ 18 & 06:52 & 51165.2891 & 0.018 & $ 101 \pm 78$ \\
      \hline
    \end{tabular}
    \begin{list}{}{}
    \item[$^\mathrm{a}$] Using the ephemeris of Navarro et al.
      (\cite{navarro}):\\ $T_0$=MJD$_\mathrm{bar}$ 49150.6086,
      $P_b$=2.0288461 days.
    \end{list}
  \end{table}

  The two radial velocities were fitted to a circular orbit using the
  ephemeris of Navarro et al.\ (\cite{navarro}).  We find a systemic
  velocity of $\gamma=-57\pm52$~km~s$^{-1}$ and a radial-velocity
  amplitude of $K_\mathrm{WD}=159\pm50$~km~s$^{-1}$. Naturally, this is a
  perfect fit, because we use two data points to determine two unknowns,
  i.e., we have no degrees of freedom and thus no information about the
  correctness of the fit.  This makes the systemic velocity and the
  velocity amplitude very dependent on any error in the determinations
  of the radial velocities.  We note, though, that possible systematic
  errors in the radial-velocity determinations have less effect on the
  velocity amplitude than on the systemic velocity, because, to first
  order, such errors should cancel out when determining the velocity
  amplitude, while being compounded in the systemic velocity.

  From the radial velocity amplitude of the white dwarf and the pulsar,
  $21.3243\pm0.0001$~km~s$^{-1}$, (Navarro et al.\ \cite{navarro}), it
  follows that $q\equiv M_\mathrm{PSR}/M_\mathrm{WD} =
  7.5\pm2.4$. 

  \section{Mass, age and distance}
  With our effective temperature and estimate of the surface gravity, we
  can use evolutionary models to estimate the mass and age of the white
  dwarf as well as the distance of the system.

  \subsection{White dwarf mass}
  The weak constraint that our spectroscopic observations place on the
  surface gravity translate into a similarly weak constraint on the
  mass. Using the models of Rohrmann et al.\ (\cite{rohrmann_b}), which
  span the range of 0.148--0.406~ $\mathrm{M}_\odot$, we find a white
  dwarf mass of $M_\mathrm{WD}=0.21^{+0.17}_{-0.04}~\mathrm{M}_\odot$.

  Similar values are found using the white dwarf evolution models of
  Driebe et al.\ (\cite{driebe_a}). We stress, though, the uncertainty:
  the upper 2$\sigma$ error on the surface gravity is $\log g=8$ (see Fig.~\ref{tefflogg}), which
  corresponds to a mass of about 0.6~$\mathrm{M}_\odot$ which is above
  the range expected for a Helium--core white dwarf.

  We can also use the mass ratio, determined above, to estimate the mass
  of the white dwarf.  If we assume a pulsar mass of $1.48\pm0.08\
  \mathrm{M}_\odot$, the weighed average of pulsar masses from
  Table~\ref{bmsp} (excluding that of PSR~J0218+4232), this results in a white dwarf mass of
  $M_\mathrm{WD}=0.20\pm0.06\ \mathrm{M}_\odot$, in line with what we
  determine from the surface gravity. The pulsar mass, as predicted by
  our observed values, is $M_\mathrm{PSR}=1.6^{+2.2}_{-0.7}~\mathrm{M}_\odot$.

  \subsection{Distance and age}
  To compare the white dwarf cooling age with the characteristic age of
  the pulsar we need to know the white dwarf's mass and temperature, as
  well as the distance to the system. The latter can be found from
  comparison with the white dwarf's absolute magnitude and our
  photometry, but for this we, again, need to know the mass of the white
  dwarf.

  Since our constraint on the white dwarf mass is weak, we invert the
  procedure. For a given white dwarf model and a given mass, we compare our
  temperature with the predictions from the model to infer a cooling age
  and use the luminosity or the absolute visual magnitude to find a distance.

\begin{figure}
    \resizebox{\hsize}{!}{\includegraphics[angle=270]{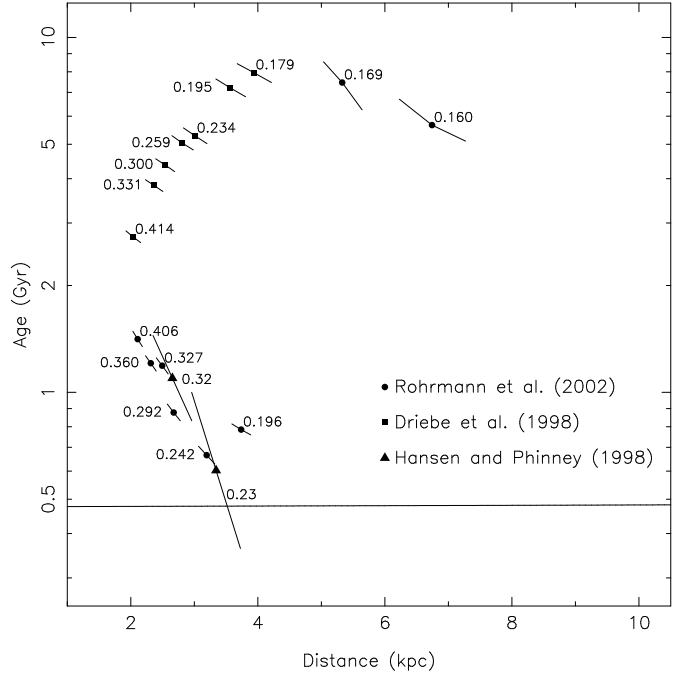}}
    \caption{The white dwarf cooling age as a function of distance.
      For each model, we determine the distance and cooling age for
      our observed temperature and magnitude. Since this, through the
      model, creates a correlation between the
      distance and cooling age, we use diagonal error bars. The two
      Hansen and Phinney models have larger errors, because we had to
      use our more uncertain I--band magnitude. The dots are the
      Rohrmann models, the squares the Driebe models and the triangles
      the Hansen and Phinney models. The solid curve denotes the
      characteristic pulsar age.}
    \label{ages}
  \end{figure}

  By comparison of our temperature with the cooling models of Hansen \& Phinney
  (\cite{hansen_a}) and the evolutionary models by Driebe et al.\ (\cite{driebe_a}) and Rohrmann et
  al.\ (\cite{rohrmann_b}) we estimate the cooling age and the distance
  of the white dwarf as a function of its mass: we show the result in
  Fig.~\ref{ages}. Each point represents a model with a certain
  mass. The errors in the distance and the age are caused by the
  uncertainty in the effective temperature of the white dwarf and, to
  a lesser extent, the uncertainty in the observed magnitudes. The
  errors are depicted by diagonal bars instead of horizontal and
  vertical ones, because the uncertainty in the distance is correlated with
  the uncertainty in the age, as both depend on the white dwarf
  temperature. The horizontal line depicts the characteristic
  age of the pulsar.

  Figure~\ref{ages} shows that, for this temperature and observed
  magnitude, there is a clear distinction between the models
  predicting thin Hydrogen envelopes and those predicting thick
  envelopes. All models 
  by Driebe et al.\ (\cite{driebe_a}) and the two lowest mass models 
  by Rohrmann et al.\ (\cite{rohrmann_b}) have thick Hydrogen
  envelopes and predict white dwarf cooling ages that are much
  older than the characteristic age of the pulsar. The Hansen \& Phinney
  (\cite{hansen_a}) models and the Rohrmann et al. (\cite{rohrmann_b})
  models with masses above $0.18\ \mathrm{M}_\odot$ have thin Hydrogen
  envelopes and hence predict younger white dwarfs at the temperature
  found for the companion of PSR~J0218+4232. If the age of the pulsar is
  indeed roughly equal to the characteristic age, then the white dwarf
  companion must have a thin Hydrogen envelope.

  The second thing to note is the close agreement with which the
  different groups predict the distance for a given mass. Since the
  luminosity we observe depends on the temperature, distance and radius
  of the white dwarf, similar distances at a given temperature indicate
  similar radii. Apparently, the radius is set by the degenerate core,
  the thickness of the Hydrogen envelope having little influence at this
  temperature.

  Returning to the comparison between the characteristic age of the
  pulsar and the cooling age of the white dwarf we can conclude that
  Helium--core white dwarfs with thin Hydrogen envelopes predict ages
  that are most similar, but a little older, to the characteristic age
  of the pulsar. The best agreement between both ages is found for
  white dwarfs with masses between 0.19 to $0.30\ \mathrm{M}_\odot$,
  corresponding to distances of 2.5 to 4 kpc and cooling ages around 0.7~Gyr. 

  \begin{table*}
    \caption[]{White dwarf companions to millisecond pulsars. The mass
      and temperature of the white dwarf in these systems is known
      accurately. Listed here are the orbital period, the mass of
      the white dwarf and of the pulsar, the temperature of the
      white dwarf and the characteristic age of the pulsar.}
    \label{bmsp}
    \begin{tabular}{lcccccc}
      \hline
      Pulsar Name & $P_\mathrm{b}$ (d) & $M_\mathrm{WD}$ ($\mathrm{M}_\odot$) &
      $M_\mathrm{PSR}$ ($\mathrm{M}_\odot$) & $T_\mathrm{eff}$ (K) &
      $\tau_\mathrm{char}$
      (Gyr) & References \\
      \hline
      PSR J1012+5307 &  0.605 & $0.16\pm0.02$ &
      $1.52\pm0.21^\mathrm{a}$/$1.64\pm0.22$ & $8550\pm25$/$8670\pm300$ & $8.9\pm1.9$ &
      1$^\mathrm{b}$, 2$^\mathrm{cd}$, 3$^\mathrm{cd}$ \\
      PSR J0437$-$4715 & 5.741 & $0.236\pm0.017$ & $1.58\pm0.18$ &
      $\sim4750$ & $4.9\pm0.1$ & 4$^\mathrm{bc}$, 5$^\mathrm{d}$ \\
      PSR B1855+09   & 12.327 & $0.248\pm0.011$ & $1.57\pm0.11$ &
      $4800\pm800$ & $4.9\pm0.2$ & 6$^\mathrm{bc}$, 7$^\mathrm{c}$, 8$^\mathrm{d}$\\[0.8ex]
      PSR J0218+4232 & $2.029$ & $0.21^{+0.17}_{-0.04}$ &
      $1.6^{+2.2}_{-0.7}$ & $8060\pm150$ & $\sim0.46$ &
      9$^\mathrm{b}$, 10$^\mathrm{cd}$\\
      \hline
    \end{tabular}\\
    References: (1) Lange et al.\ (\cite{lange}), (2) van Kerkwijk et
    al.\ (\cite{kerkwijk_a}), (3) Callanan et al.\ (\cite{callanan}),
    (4) van Straten et al.  (\cite{straten}),\\ (5) Danziger et al.\
    (\cite{danziger}), (6) Kaspi et al.\ (\cite{kaspi}), (7) Nice et al. (\cite{nice}),
    (8) van Kerkwijk et al. (\cite{kerkwijk_b}), (9) Navarro et
    al. (\cite{navarro})\\ and (10) this paper.\\
    $^\mathrm{a}$ Using the white dwarf mass and a mass-ratio of
    $9.5\pm0.5$ (Unpublished work).\\
    $^\mathrm{b}$ Timing observations.\\
    $^\mathrm{c}$ Mass determination.\\
    $^\mathrm{d}$ Temperature determination.\\
  \end{table*}

  \section{Discussion and conclusions}
  We have identified the companion of PSR~J0218+4232 and presented
  photometry and spectroscopy. Our spectra show that the companion has a
  Hydrogen atmosphere and from comparison with Hydrogen atmosphere
  models we find that the object has temperature of 
  $T_\mathrm{eff}=8060\pm150$~K and a surface gravity of 
  $\log g=6.9\pm0.7$ cm s$^{-1}$. 

  This temperature and gravity would imply a mass of
  $M_\mathrm{WD}=0.21^{+0.17}_{-0.04}~\mathrm{M}_\odot$, in line with
  what one would expect for a Helium--core white dwarf.
  Similar white dwarf masses, 0.19 to $0.30\ \mathrm{M}_\odot$, are
  found when comparing the predicted cooling age of the white dwarf
  with the characteristic age of the pulsar.

  Furthermore, the comparison between the cooling age and the
  characteristic age shows that this white dwarf companion is likely 
  to have a thin Hydrogen envelopes, in agreement with the prediction of
  Althaus et al.\ (\cite{althaus}), that the transition
  between thick and thin Hydrogen envelopes is situated around a white
  dwarf mass of $\sim0.18\ \mathrm{M}_\odot$. 

  Thin Hydrogen envelopes are also required to match the white dwarf
  cooling ages with characteristic pulsar ages in the systems PSR
  B1855+09 and PSR J0437$-$4715 (Table~\ref{bmsp}). For PSR
  J1012+5307, several authors (Alberts et al.\ \cite{alberts}, Driebe
  et al.\ \cite{driebe_a}, Althaus et al.\ \cite{althaus}) have
  argued that a thick Hydrogen layer is needed to explain its old age
  and high temperature. 
  Our mass determination is to weak to put a direct observational constraint
  on the location of the transition.  On the basis of the relation
  between orbital period and mass, however, we expect that the mass of
  PSR J0218+4232 is in between those of PSR J1012+5307 and PSR
  J0437$-$4715.  We can conclude that for whatever mass corresponds to
  an orbital period of 2 days, thin Hydrogen layers are needed.

  This empirical constraint may be improved from observations of the white
  dwarf companions of millisecond pulsars \object{PSR~J0034$-$0534} and
  \object{PSR~J0613$-$0200}, as these systems have orbital periods, 1.6
  and 1.1 d respectively, in between those of PSR~J0218+4232 and
  PSR~J1012+5307. 

  The companion of PSR~J0034$-$0534 has been detected by Lundgren et al.\
  (\cite{lundgren_b}) who place an upper limit on the temperature of
  3800~K. This limit depends on the distance, which was inferred from
  the dispersion measure. Nevertheless, with a characteristic pulsar
  age of 5.9~Gyr, this white dwarf would likely need a thin Hydrogen envelope
  to have cooled to such a low temperature.

  So far, the only observed system for which a thick Hydrogen envelope is
  required, is PSR J1012+5307. We should keep in mind, however, the
  possibility that this pulsar is actually very young, i.e., that the
  assumption $P_0 \ll P$ does not hold. If this were the case then
  perhaps no white dwarf would need thick Hydrogen layers. 

  This can be verified observationally by measuring the temperature of
  the white dwarf companion of \object{PSR~J0751+1807}.  This binary
  has an orbital period of only 6.3~hours, shorter than that of
  PSR~J1012+5307, and hence the white dwarf should have even lower
  mass.  Optical observations by Lundgren et 
  al.\ (\cite{lundgren_a}) have not detected an optical counterpart
  down to $V=23.5$, indicating that the white dwarf companion is
  rather cool ($T\la7000~$K for $M<0.2~\mathrm{M}_\odot$ and
  $d\simeq2~$kpc).  It would need a thin Hydrogen layer to have
  cooled to such a low temperature within the characteristic age of the
  pulsar (8~Gyr). Identification of this counterpart and a measurement
  of its temperature could decide whether it has a thin or a thick Hydrogen
  layer.

  \begin{acknowledgements}
    C.G.B. is indebted to F. Hulleman for help with the photometry.  We
    thank P. Bergeron, A. Serenelli and B. Stappers for very useful
    discussions about white dwarf atmospheres, white dwarf interiors and radio
    observations, respectively.  The observations reported here were
    obtained at the W.M. Keck Observatory, which is operated by the
    California Association for Research in Astronomy.  The reduction was
    done using the Munich Image Data Analysis System, which is developed
    and maintained by the European Southern Observatory.  This research
    made use of the SIMBAD database.  M.H.v.K. acknowledges support of a
    fellowship of the Royal Netherlands Academy of Arts and Sciences,
    and S.R.K. of grants from NSF and NASA.
  \end{acknowledgements}


\begin{thebibliography}{}
  \bibitem[1996]{alberts} Alberts, F., Savonije, G. J., van den Heuvel, E. P. J.,
    Pols, O. R. 1996, Nature 380, 676
  \bibitem[2001]{althaus} Althaus, L. G., Serenelli, A. M.,
    Benvenuto, O. G. 2001, MNRAS 324, 617
  \bibitem[1995]{bergeron_a} Bergeron, P., Wesemael, F., Beauchamp, A. 1995, PASP 107, 1047
  \bibitem[1990]{bessell} Bessell, M. S. 1990, PASP 102, 1181
  \bibitem[1995]{bohlin} Bohlin, R. C., Colina, L., Finley, D. S. 1995, AJ 110, 1316
  \bibitem[1998]{callanan} Callanan, P. J., Garnavich, P. M., Koester, D. 1998, MNRAS 298, 207
  \bibitem[1994]{camilo} Camilo, F., Thorsett, S. E., Kulkarni S. R. 1994, ApJ 421, L15
  \bibitem[2002]{cordes} Cordes, J. M. \& Lazio, T. J. W. 2002, preprint
    [\texttt{astro-ph/0207156}]
  \bibitem[1993]{danziger} Danziger, I. J., Baade, D., Della Valle, M. 1993, A\&A 276, 382
  \bibitem[1998]{driebe_a} Driebe, T., Sch\"onberger, D., Bl\"ocker, T., Herwig, F. 1998,
    A\&A 339, 123
  \bibitem[1999]{driebe_b} Driebe, T., Bl\"ocker, T., Sch\"onberger,
    D., Herwig, F. 1999, A\&A 350, 89
  \bibitem[1998a]{hansen_a} Hansen, B. M. S. \& Phinney, E. S. 1998a, MNRAS 294, 557
  \bibitem[1986]{horne} Horne, K. 1986, PASP 98, 609
  \bibitem[1987]{joss} Joss, P. C., Rappaport, S., Lewis, W. 1987,
    ApJ 319, 180
  \bibitem[1994]{kaspi} Kaspi, V. M., Taylor, J. H., Ryba, M. F. 1994, ApJ 428, 713
  \bibitem[1992]{landolt} Landolt, A. U. 1992, AJ 104, 340
  \bibitem[2001]{lange} Lange, Ch., Camilo, F., Wex, N. et al. 2001, MNRAS 326, 274
    J.M. 1995, ApJ 453, 419
  \bibitem[1996a]{lundgren_a} Lundgren, S. C., Cordes, J. M., Foster, R. S., Wolszczan, A., Camilo, F. 1996a, ApJ 458, L33
  \bibitem[1996b]{lundgren_b} Lundgren, S. C., Foster, R. S., Camilo,
    F. 1996b, Johnston, S., Walker, M. A., Bailes, M. (eds.) Pulsars:
    Problems \& Progress. ASP Conf. Ser. Vol. 105, p. 497
  \bibitem[1998]{lyne} Lyne, A. G. \& Smith, F. G. 1998 Pulsar
    Astronomy (Cambridge: Cambrigde Univ. Press)
  \bibitem[1998]{monet} Monet, D., et al. 1998, USNO-A2.0 (Washington DC: US Naval
    Observatory)
  \bibitem[1995]{navarro} Navarro, J., de Bruyn, A. G., Frail, D. A., Kulkarni, S. R.,
    Lyne, A. G. 1995, ApJ 455, L55
  \bibitem[1995]{nicastro} Nicastro, L., Lyne, A. G., Lorimer, D. R. et al. 1995, MNRAS 273, L68
  \bibitem[2002]{nice} Nice, D. J., Splaver, E. M., Stairs,
    I. H. 2002, preprint [\texttt{astro-ph/0210637}]
  \bibitem[1994]{odonnell} O'Donnell, J. E. 1994, ApJ 422, 158
  \bibitem[1995]{oke} Oke, J. B., Cohen, J. G., Carr, M., et al. 1995,
    PASP 107, 375
  \bibitem[1995]{rappaport} Rappaport, S., Podsiadlowski, Ph., Joss,
    P. C., Di Stefano, R., Han, Z. 1995, MNRAS 273, 731
  \bibitem[1996]{reid} Reid, I. N. 1996, AJ 111, 2000
  \bibitem[2001]{rohrmann_a} Rohrmann, R. D. 2001, MNRAS 323, 699
  \bibitem[2002]{rohrmann_b} Rohrmann, R. D., Serenelli, A. M., Althaus, L. G., Benvenuto,
    O. G. 2002, MNRAS, 335, 499
  \bibitem[2000]{sarna_a} Sarna, M. J., Ergma, E., Antipova, J. 2000, MNRAS 316, 84
  \bibitem[2001]{sarna_b} Sarna, M. J., Ergma, E.,
    Ger$\breve{\mathrm{s}}$kevit$\breve{\mathrm{s}}$, J. 2001, AN 322, 405
  \bibitem[1998]{schlegel} Schlegel, D. J., Finkbeiner, D. P., Davis, M. 1998, ApJ 500, 525
  \bibitem[2000]{schoenberger} Sch\"onberger, D., Driebe, T.,
    Bl\"ocker, T. 2000, A\&A 356, 929
  \bibitem[2001]{serenelli} Serenelli, A. M., Althaus, L. G.,
    Rohrmann, R. D., Benvenuto, O. G. 2001, MNRAS 325, 607
  \bibitem[1970]{shklovskii} Shklovskii, I. S. 1970, Soviet Astron., 13, 562
  \bibitem[1987]{stetson_a} Stetson, P. B. 1987, PASP 99, 191
  \bibitem[2000]{stetson_b} Stetson, P. B. 2000, PASP 112, 925
  \bibitem[1999]{tauris} Tauris, T. M. \& Savonije, G.J. 1999, A\&A 350, 928
  \bibitem[1993]{taylor} Taylor, J. H. \& Cordes, J. M. 1993 ApJ 411, 674
  \bibitem[1999]{thorsett} Thorsett, S. E. \& Chakrabarty, D.  1999, ApJ 512, 288
  \bibitem[1996]{kerkwijk_a} van Kerkwijk, M. H., Bergeron, P., Kulkarni, S. R. 1996, ApJ 467, L89
  \bibitem[2000]{kerkwijk_b} van Kerkwijk, M. H., Bell, J. F., Kaspi, V. M., Kulkarni, S. R. 2000 ApJ 530, L37
  \bibitem[2001]{straten} van Straten, W., Bailes, M., Britton, M. et al. 2001, Nature 412, 158
  \bibitem[1975]{webbink} Webbink, R. F. 1975 MNRAS 171, 555
  \end{thebibliography}
\end{document}